\documentclass[prl,twocolumn,aps,letterpaper, preprintnumbers,superscriptaddress]{revtex4-2}

\usepackage{graphicx}
\usepackage{dcolumn}
\usepackage{bm}

\usepackage{amsthm,amsmath,amssymb}
\usepackage{mathrsfs}
\usepackage{appendix}
\usepackage{amstext} 
\usepackage{url}
\usepackage{float} 
\usepackage{soul}
\usepackage[usenames,dvipsnames]{color}
\usepackage[colorlinks=true,citecolor=blue,urlcolor=black]{hyperref}
\usepackage{dcolumn}
\usepackage{booktabs}
\usepackage[linesnumbered,algoruled,boxed,lined]{algorithm2e} 
\usepackage{siunitx}
\usepackage{makecell}
\usepackage{ragged2e}

\newcommand{\zy}{\textcolor{black}}
\newcommand{\lz}{\textcolor{black}}

\begin{document}
\title{Post-Measurement Pairing 
Quantum Key Distribution\\ with Local Optical Frequency Standard}
\author{Chengfang Ge}
\affiliation{Beijing Academy of Quantum Information Sciences, Beijing 100193, China} 
\affiliation{Institute of Physics, Chinese Academy of Sciences, Beijing 100190, China}
\affiliation{University of Chinese Academy of Sciences, Beijing 101408, China}
\author {Lai Zhou}\email{zhoulai@baqis.ac.cn}

\affiliation{Beijing Academy of Quantum Information Sciences, Beijing 100193, China} 
\author{Jinping Lin}
\affiliation{Beijing Academy of Quantum Information Sciences, Beijing 100193, China} 
\author{Hua-Lei Yin}
\affiliation{Beijing Academy of Quantum Information Sciences, Beijing 100193, China} 
\affiliation{Department of Physics and Beijing Key Laboratory of Opto-Electronic Functional Materials and Micro-Nano Devices, Key Laboratory of Quantum State Construction and Manipulation (Ministry of Education), Renmin University of China, Beijing 100872, China}
\author{Qiang Zeng}
\affiliation{Beijing Academy of Quantum Information Sciences, Beijing 100193, China} 
\author{Zhiliang~Yuan}
\affiliation{Beijing Academy of Quantum Information Sciences, Beijing 100193, China} 

\date{\today}

\begin{abstract}
The idea of post-measurement coincidence pairing simplifies substantially long-distance, repeater-like quantum key distribution (QKD) by eliminating the need for tracking the differential phase of the users' lasers. However, optical frequency tracking remains necessary and can become  a severe burden in future deployment of multi-node quantum networks.  Here, we resolve this problem by referencing each user's laser to an absolute frequency standard and demonstrate a practical post-measurement pairing QKD with excellent long-term stability. We confirm the setup's repeater-like behavior and achieve a finite-size secure key rate (SKR) of 15.94~bit/s over 504~km fiber, which overcomes the absolute repeaterless bound by 1.28~times. Over a fiber length 100 km, the setup delivers an impressive SKR of 285.68 kbit/s.  Our work paves the way towards an efficient muti-user quantum network with the local frequency standard.
\end{abstract} 
 
%
\maketitle
\textit{Introduction.}---
Quantum key distribution (QKD) 
has emerged as an important solution to mitigate the threats posed by quantum computers to secure data communication~\cite{dynes2019cambridge,chen2021integrated}. Remarkable achievements  have been made in  
secure key rates (SKRs)~\cite{yuan201810,li2023high
}, transmission distance~\cite{PhysRevLett.121.190502} and miniaturization~\cite{
paraiso2021photonic,sax2023high
} while preserving
implementation security~\cite{xu2020secure, zapatero2024implementation}. However, fiber loss limits severely the performance of  long-haul quantum transmission~\cite{pirandola17}. 
While the concept of quantum repeaters~\cite{briegel1998quantum,duan2001long} 
promises high-performance  communication over arbitrarily long distances, the technology is still in the development phase, with the longest reaches of just dozens of kilometers~\cite{
van2022entangling}.

With available technologies, twin-field (TF)~\cite{Lucamarini2018} and post-measurement pairing (PMP)~\cite{zeng2022mode,xie2022breaking} QKD protocols have been verified to deliver repeater-like secure key rates (SKRs)~\cite{pittaluga21,wang22,zhou2023quantum, zhong2021proof, 
PhysRevLett.130.250801,liu2023experimental}. 
Their security relies upon interference at the intermediate measurement node (Charlie) between the coherent fields prepared by the remote communication users. 
Successful operation of TF-QKD requires stringent optical phase tracking between the users' lasers, while PMP-QKD relaxes it to simpler frequency tracking~\cite{avesani2023long}
as demonstrated by either real-time monitoring the optical frequencies~\cite{zhu2022experimental} or locally stabilizing each user's laser to an ultra-stable cavity\lz{~\cite{PhysRevLett.130.250801}}. 
However, the solutions remain burdensome, and could rapidly become intractable in a network scenario~\cite{wei2020high}, where the optical frequencies at multiple nodes have to be kept in alignment. 

In the telecommunication C-band, the acetylene ($^{13}\mathrm{C}_2\mathrm{H}_2$)  $\mathrm{P_{16}}$ $(\nu_1 + \nu_3)$ transition line at $\lambda \approx 1542.384$~nm is an absolute frequency reference that has been officially recommended by the CIPM~\cite{fre_standard}. This standard offers a relative (frequency) uncertainty of $2.6 \times 10^{-11}$ (5~kHz), which \zy{would produce phase instability} on the same order of magnitude as fiber length fluctuations found in long-distance links~\cite{Lucamarini2018}, and presents itself as an attractive path to unify optical frequencies across a quantum network and completely eliminate the burden of frequency tracking. 

Here, we employ independent lasers stabilized to the absolute frequency standard and demonstrate the first repeater-like QKD without optical frequency tracking. 
Despite the phase instability of the compact lasers, it remains possible to perform time-bin two-photon interference using bin separation of the order of 100~$\mu$s for efficient and low-error post-measurement coincidence pairing. 
At a fiber length of 100~km, our PMP-QKD setup delivers an impressive SKR exceeding 285.68~kbit/s, which compares favourably with state-of-the-art measurement-device-independent~\cite{lo2012measurement} (MDI) QKD systems~\cite{wei2020high,woodward2021gigahertz,zhu2022experimental}.  Its repeater-like behavior is verified by observation of the SKR's square-root dependence on the fiber distance. At 504.46~km, we obtain a finite-size SKR of \lz{15.94}~bit/s, which exceeds the absolute repeater-less bound  by a factor of \lz{1.28}.

\begin{figure*}[t]
\centering
\includegraphics[width=18cm]
{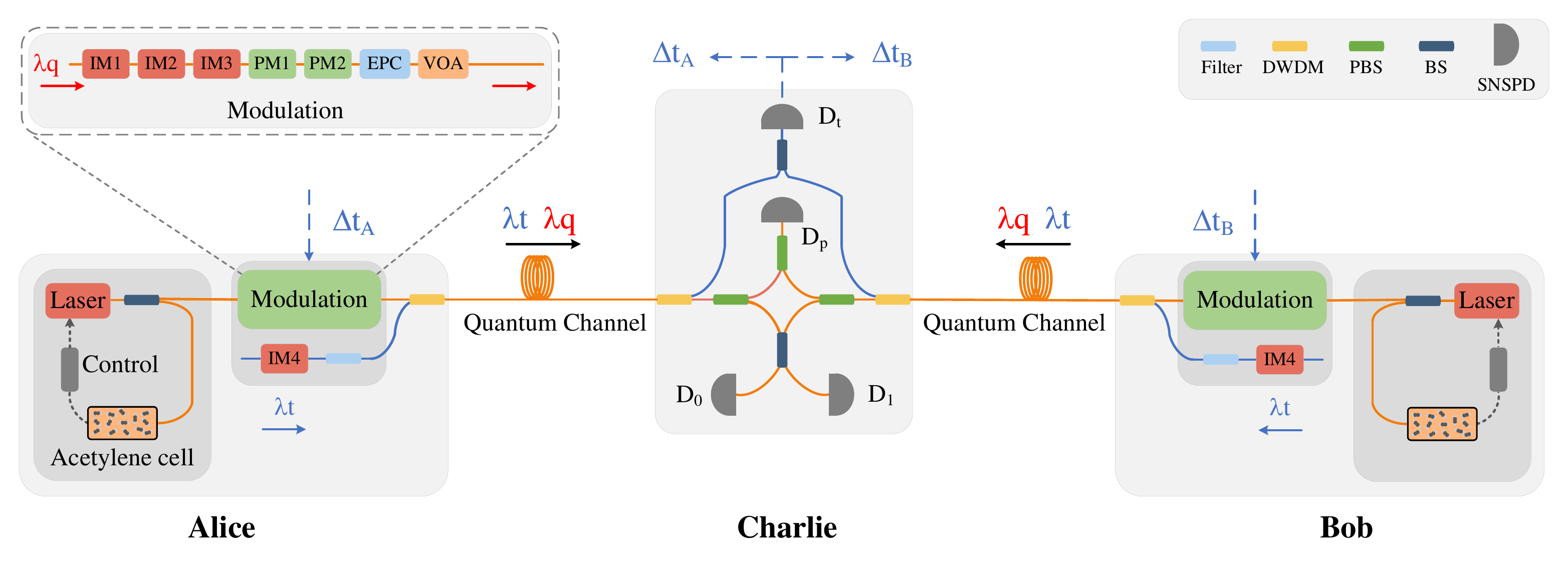}	
\caption{\textbf{Experimental setup.} 
Each user (Alice or Bob) 
owns an independent laser whose wavelength is referenced to the saturation absorption line \zy{(1542.384~nm)} of acetylene. The modulation unit encodes the incoming continuous-wave laser signal into weak coherent pulses carrying quantum information. Quantum signals ($\lambda_q$) are transmitted to the measurement node Charlie, together with 10 MHz timing pulses ($\lambda_t$) generated by intensity modulating a DFB laser (not shown). The interference result of the quantum signals is registered by $D_0$ and $D_1$. Count rates in $D_p$ and $D_t$ serve as error signals, allowing the compensation of polarization and time fluctuations in the quantum channel, respectively. PM: phase modulator; EPC: electrically driven polarization controller;  VOA: variable optical attenuator; DWDM: dense wavelength division multiplexer; PBS: polarization beam splitter; BS: beam splitter; SNSPD: superconducting nanowire single photon detector.}
 \label{fig_experiment_setup}
\end{figure*}

\textit{Experimental Setup.}---
The experimental setup (Fig.~\ref{fig_experiment_setup}) consists of three modules: the communication users Alice and Bob and the intermediate measurement node Charlie. Each user owns a compact rack-mountable continuous-wave laser with its wavelength locked locally to the intrinsic transition of 1542.38~nm of acetylene molecules. The absolute reference ensures Alice and Bob's lasers' suitability for encoding quantum signals and sufficiently stable first-order interference.
The laser signal ($\lambda_q$) is first carved into a train of pulses with a width of 300~ps clocked at 1~ns intervals. These pulses are then modulated using a cascade of intensity modulators (IMs) and phase modulators (PMs), contained in the modulation unit (inset, Fig.~\ref{fig_experiment_setup}), to generate a waveform pattern that meets the requirement to implement the version of the PMP-QKD protocol, which was originally referred to as asynchronous MDI-QKD~\cite{xie2022breaking,PhysRevLett.130.250801}.
Each user pre-compensates the polarization rotation of their quantum signal by the fiber channel using an electrically driven polarization controller (EPC). Before entering the quantum channel, the encoded pulses are attenuated to the single photon level using a variable optical attenuator (VOA). 
The quantum transmission maintains 100~\% duty cycle.

Synchronization between Alice and Bob's modulation units is provided an 50~MHz electrical clock distributed by Charlie, as indicated by blue dash lines in Fig.~\ref{fig_experiment_setup}.
To mark the arrival of their quantum pulses, each user has a continuous-wave distributed feedback (DFB) laser of a central wavelength of 1535.04~nm 
that is carved by an intensity modulator to produce  10~MHz pulses of 15~ns width. This wavelength signal ($\lambda_t$) is spectrally cleaned by a 100~GHz bandpass filter before being merged into the quantum channel using a dense wavelength division multiplexer (DWDM). 
The quantum channel is composed of ultra-low-loss fiber spools (G654.C ULL), which have average loss coefficients ranging from 0.159~dB~km$^{-1}$ to 0.168~dB~km$^{-1}$.

Upon receiving the incoming signals,  Charlie uses his DWDM filters to separate out the $\lambda_t$ signals and routes them to a superconducting nanowire single-photon detector (SNSPD) $D_t$. Through histogramming, the detector monitors the arrival times of each user's $\lambda_t$ signals, and Charlie then adjusts accordingly the delays ($\Delta t_A$ and $\Delta t_B$) of his clock signals transmitted to Alice and Bob.  This automated compensation routine ensures optimal temporal alignment and thus high-visibility interference between Alice and Bob's quantum pulses. Note that the measurement by $D_t$ is incoherent and therefore Alice and Bob's $\lambda_t$ lasers do not require frequency locking.

Charlie uses a polarization beam splitter (PBS) at each incoming quantum path so as to maintain an identical polarization of photons at his interfering 50/50 beam splitter. Reflection ports of the PBS's are combined and detected by detector $D_p$, the count rate of which is kept minimal by adjusting Alice and Bob's EPC's throughout each QKD session. 
After mitigating the time and polarization drifts, the quantum pulses  ($\lambda_q$)  interference at Charlie's 50/50 beam splitter (BS), the results of which are detected by detectors $D_0$ and $D_1$ with respective detection efficiencies of $71.0 \%$ and $70.5 \%$, dark count rates of 6.3~and 9.0~Hz, and a time jitter of approximately 40~ps. Details about the system loss and detectors are summarized in Appendix B.

\textit{{Protocol}.}---We adopt the PMP-QKD protocol~\cite{xie2022breaking} with three different pulse intensities ($\mu,\nu,o$), where $\mu>\nu>o=0$. 
Alice (Bob) encodes each pulse with one of the three intensities, and a phase value $\theta$ randomly selected from 16 values: $\theta_{a(b)} \in \{0,\pi/8,...,15\pi/8\}$.
A successful click, i.e., one and only one detector ($D_0$, $D_1$) registers a photon at the given time slot, 
is denoted as $(k_{a}|k_{b})$ if Alice used intensity $k_a$ and Bob $k_b$ ($k_a,k_b\in \{\mu,\nu,o\}$) for encoding.

We use a neighbor paring scheme~\cite{zeng2022mode} where adjacent clicks are paired together to form a coincidence if their temporal separation is less than the maximal pairing interval $T_c$, which is pre-set so that the $\boldsymbol{X}$-basis quantum bit error rate (QBER) is kept at an acceptable level. To improve the pairing probability, a simple filter~\cite{PhysRevLett.130.250801} is applied to discard clicks ($\mu_a|\nu_b$) and ($\nu_a|\mu_b$) before pairing. $[k_{a}^{tot},k_{b}^{tot}]$ represents the coincidence where $k_{a}^{tot}$ ($k_{a}^{tot}$) is the combined intensity of two time bins in Alice's (Bob's) side. Coding basis for each coincidence is assigned according to the parameters: $k_{a}^{tot}$ and $k_{b}^{tot}$. For $[\mu_a,\mu_b]$,  Alice (Bob) extracts a $\boldsymbol{Z}$-basis bit 0 (1) if he sends $\mu_a$ ($\mu_b$) in the early ($e$) time bin and $o_a$ ($o_b$) in the late ($l$) time bin. Otherwise, Alice (Bob) extracts an opposite bit value. $\boldsymbol{X}$-basis bits are extracted from  those $[2\nu_a,2\nu_b]$ coincidences that Alice and Bob's phase modulations meet the condition that $\phi_{ab} = (\theta_a^l - \theta_a^e - \theta_b^l + \theta_b^e) \mod 2\pi = 0$ or $\pi$. $\boldsymbol{X}$-basis bits are used only for bounding the information leakage, while just $\boldsymbol{Z}$-basis bits contribute to the raw key, from which secure key bits are extracted after error correction and privacy amplification. The SKR considering finite size effects is calculated as,
\begin{equation}
\begin{aligned}
R=&\frac{F}{N}\left\{\underline{s}_{0}^z+\underline{s}_{11}^z\left[1-H_2\left(\overline{\phi}_{11}^z\right)\right]-\lambda_{\rm{EC}}\right.
\\ 
&-\left.\log_2\frac{2}{\varepsilon_{\rm cor}}-2\log_2\frac{2}{\varepsilon'\hat{\varepsilon}}-2\log_2\frac{1}{2\varepsilon_{\rm PA}}\right\},\label{eq_keyrate}
\end{aligned}
\end{equation}
where $F$ is the effective system clock frequency for quantum signal transmission, $N$ is the number of signal pulses Alice or Bob sent, $H_2(x)=-x\log_2x-(1-x)\log_2(1-x)$ is the binary Shannon entropy function. ${s}_{0}^z$, ${s}_{11}^z$ are the number of vacuum components and that of the single-photon pair components in the $\boldsymbol{Z}$-basis, respectively. ${\phi}_{11}^z$ is the single-photon pair phase error rate. Notations $\underline{x}$ and $\overline{x}$ denote the corresponding lower and upper bounds of parameter $x$. $\lambda_{\rm{EC}}$ is the amount of information revealed in the error correction. $\varepsilon_{\rm cor}$, $\varepsilon_{\rm PA}$, $\varepsilon'$ and $\hat{\varepsilon}$ are security coefficients about secrecy and correctness. 

\begin{figure}[t]
\centering
\includegraphics[width=\linewidth]{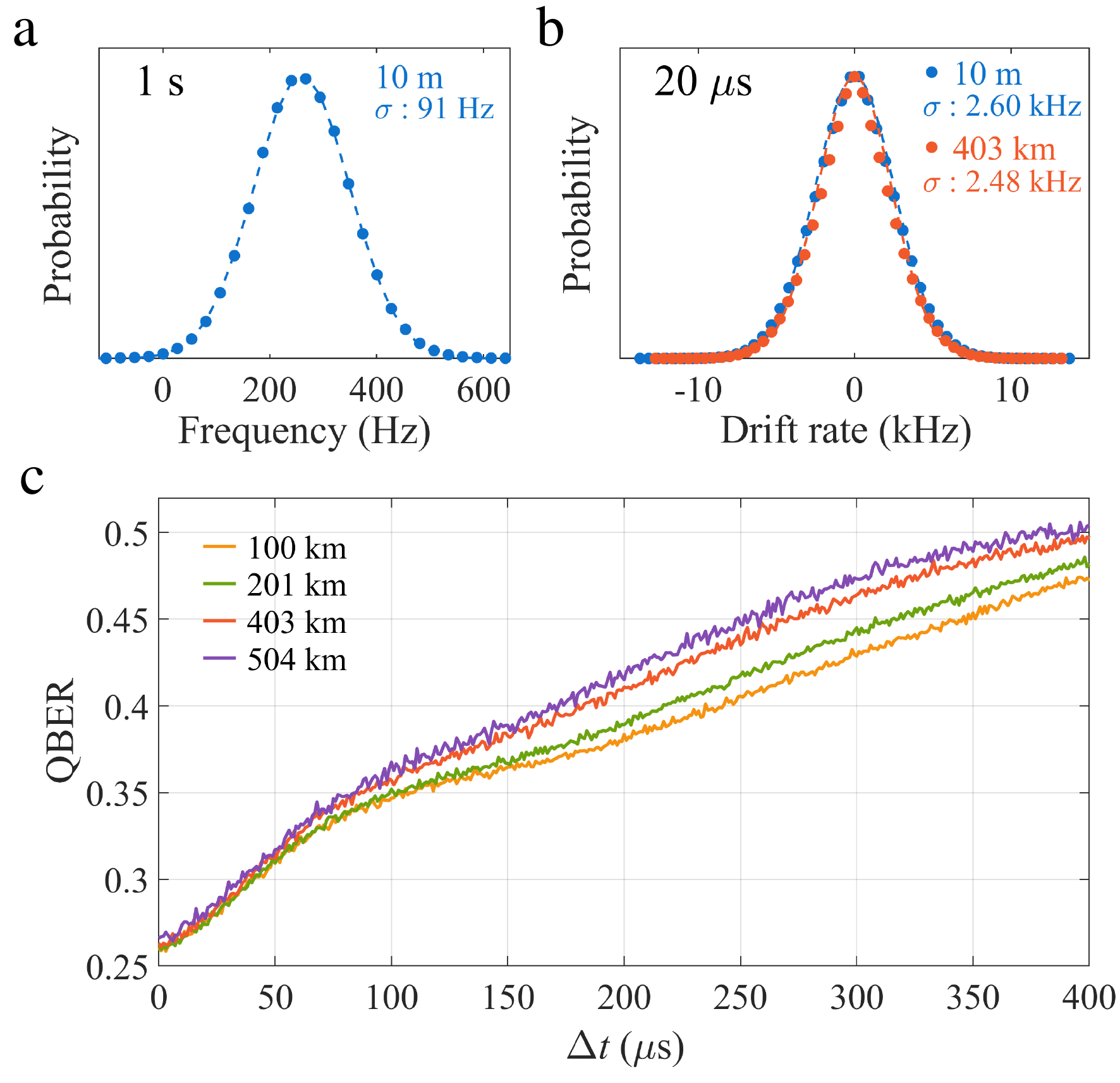}
	\caption{\textbf{Frequency/phase instability of the lasers and its impact on the $\boldsymbol{X}$-basis QBER}. \textbf{a}, Distributions of the frequency difference between the users' lasers and the standard deviation measured with 1~s intervals. \textbf{b}, Distribution of the phase drift rates measured with  20~$\mu s$ intervals without (blue dots) and with (orange dots) long fibers. \textbf{c}, $\boldsymbol{X}$-basis QBER as a function of the pairing interval for different fiber distances, ranging from $100.94$ to $504.66$ km.}\label{FRE}
\end{figure}

\textit{Experimental results.}---We first determine the beat frequency between the lasers independently locked to their own acetylene's cells using a fast photodiode and a frequency counter. 
To determine the sign of the frequency beat, an acoustic optical modulator is used to shift one laser's frequency by 80~MHz.  In the absence of long fibers, the frequency difference has a systematic offset 261~Hz and a standard deviation of \lz{91}~Hz obtained from a 60~h measurement with 1~s intervals, see Fig.~\ref{FRE}\textbf{a}.
We attribute the systematic offset to pressure variation between acetylene gas cells and nonlinearity in the piezo response of the laser frequency stabilization loop~\cite{hald2011fiber,Talvard:17}. Encouragingly, the offset is an order of magnitude smaller than fiber length fluctuation reported for a quantum channel of several hundred kilometers~\cite{Lucamarini2018,zhou2023quantum}.

To resolve faster phase drift, we use an optical power meter to sample the interference between the two lasers at an interval of $20$~$\mu$s.
As shown in Fig.~\ref{FRE}\textbf{b}, the phase drift rate distribution (blue dots) has a standard deviation of 2.60~kHz (1~Hz = 2$\pi$~rad/s), which is about 10~times larger than their systematic offset. 
Without long fiber, this phase drift rate can only be caused by the instability of the lasers due to imperfection in the locking to acetylene gas cells. 
Adding \lz{201~km} fibers after each laser, the phase drift has an even narrower distribution of 2.48~kHz (orange dots), suggesting the dominance of the laser instability over fiber fluctuation.
The narrower distribution is explained by the slow fiber drift, which happened to (partially) compensate for the systematic offset between the lasers' frequencies. 

We then investigate the impact of the instability of the acetylene-stabilized lasers on the $\boldsymbol{X}$-basis QBER. 
Here, we implement phase randomization and decoy state modulation as required by the adopted protocol~\cite{xie2022breaking}, and process the photon detection data using the pair-wise coincidence method as described in detail in~\cite{PhysRevLett.130.250801}. 
For evaluation purpose, the QBER here is extracted using the more probable $[2\mu_a,2\mu_b]$ coincidences, where Alice (Bob) sent a $\mu_a$ ($\mu_b$) pulse in either time bin of a successful post-paired coincidence. Alice and Bob extract an identical bit value from a coincidence if $\phi_{ab} = \pi$ and both detectors clicked once, or $\phi_{ab} = 0$ and the same detector clicked twice. Otherwise, they are assigned with opposite bit values. The $\boldsymbol{X}$-basis QBER has a minimum value of 0.25 because of use of weak coherent pulses~\cite{lo2012measurement}.
In Fig.~\ref{FRE}\textbf{c}, the $\boldsymbol{X}$-basis QBER is plotted as a function of time interval for different fiber distances ranging from $100.94$ to $504.66$~km. 
For all fiber lengths studied, the QBER increases significantly from 0.26 to \lz{0.32} when the pairing interval increases to 50~$\mu$s, which we attribute to the lasers' frequency instability.
As the time interval increases further, the impact of fiber fluctuation starts to show. A longer fiber has a greater impact of the fiber length fluctuation.

\begin{figure}[t]
\centering
\includegraphics[width=\linewidth]{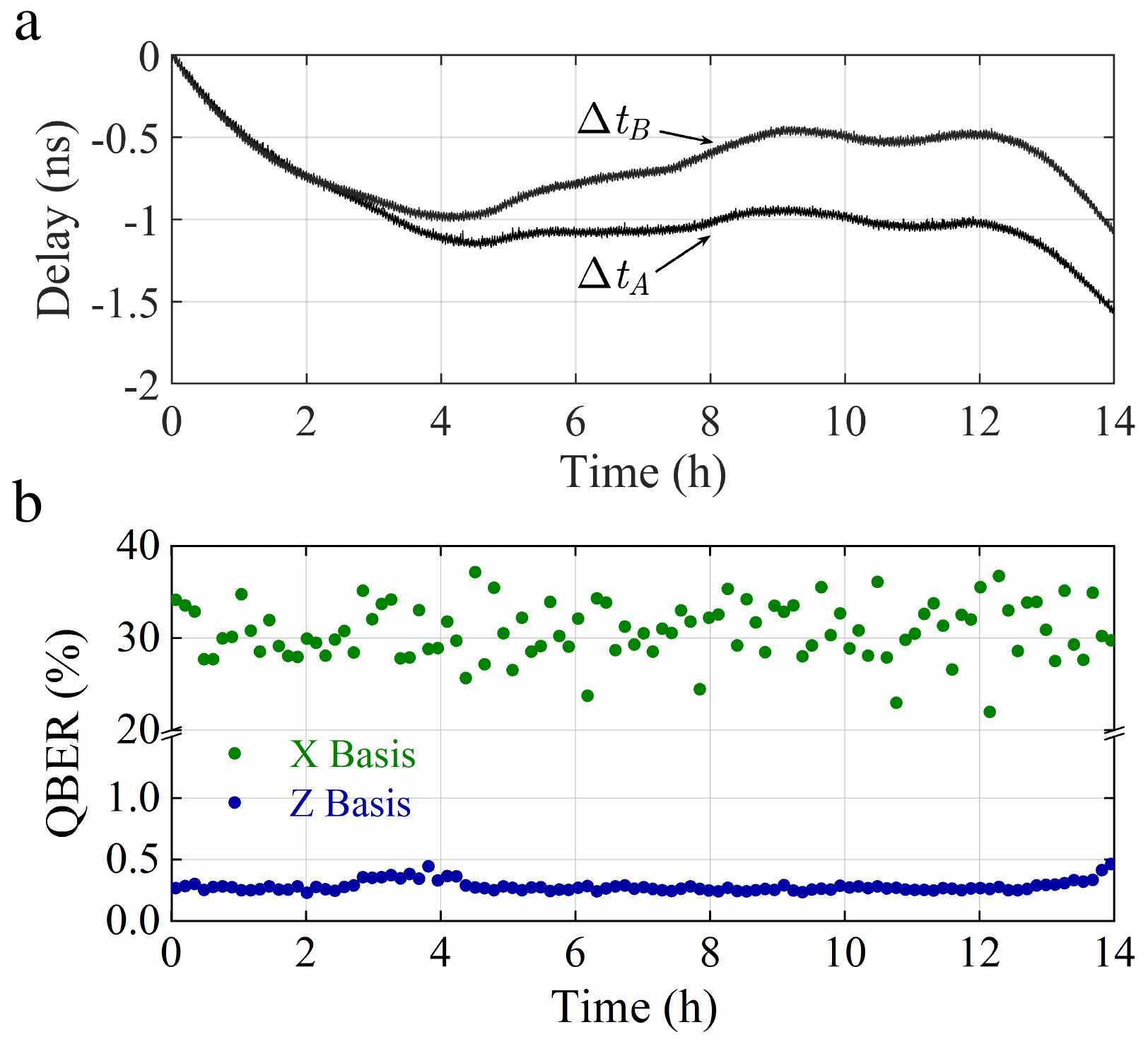}
	\caption{\textbf{Continuous operation over 504.66~km fiber link.} \textbf{a}, Clock delay $\Delta t_A$ ($\Delta t_B$) to Alice (Bob).  
 \textbf{b}, Evolution of the $\boldsymbol{X}$-basis and $\boldsymbol{Z}$-basis's QBERs. Each QBER data point is calculated from $\sim$500~s data. 
 }
 \label{HOM}
\end{figure}

Our setup runs continuously thanks to the implemented auto-compensation for temporal and polarization alignment. 
Figure~\ref{HOM} shows an example of a continuous 14~h data recorded in the experiment with 504.66~km fiber. 
Compensating of Alice and Bob's modulation delays in real time, the standard deviation of the temporal alignment is 14.9~ps (not shown), which is far below the quantum signal pulse width of 300~ps and thus ensures high-visibility interference.  
In obtaining \zy{the} $\boldsymbol{X}$-basis QBER, we cap the maximum pairing interval ($T_c$) at $77$~$\mu$s in order to achieve a balance between the pairing efficiency and the QBER. This leads to a mean pairing interval of 35.3~$\mu s$ for the dataset of this distance, \zy{with} which we obtain an $\boldsymbol{X}$-basis QBER of $(30.7 \pm 3.0)$~\%. 
\zy{The corresponding} $\boldsymbol{Z}$-basis QBER  \zy{is} just (0.24~$ \pm$ 0.04)~\%, \zy{thanks to time decoding's immunity to} phase deviation.

\begin{figure}[t]
\centering
\includegraphics[width=\linewidth]{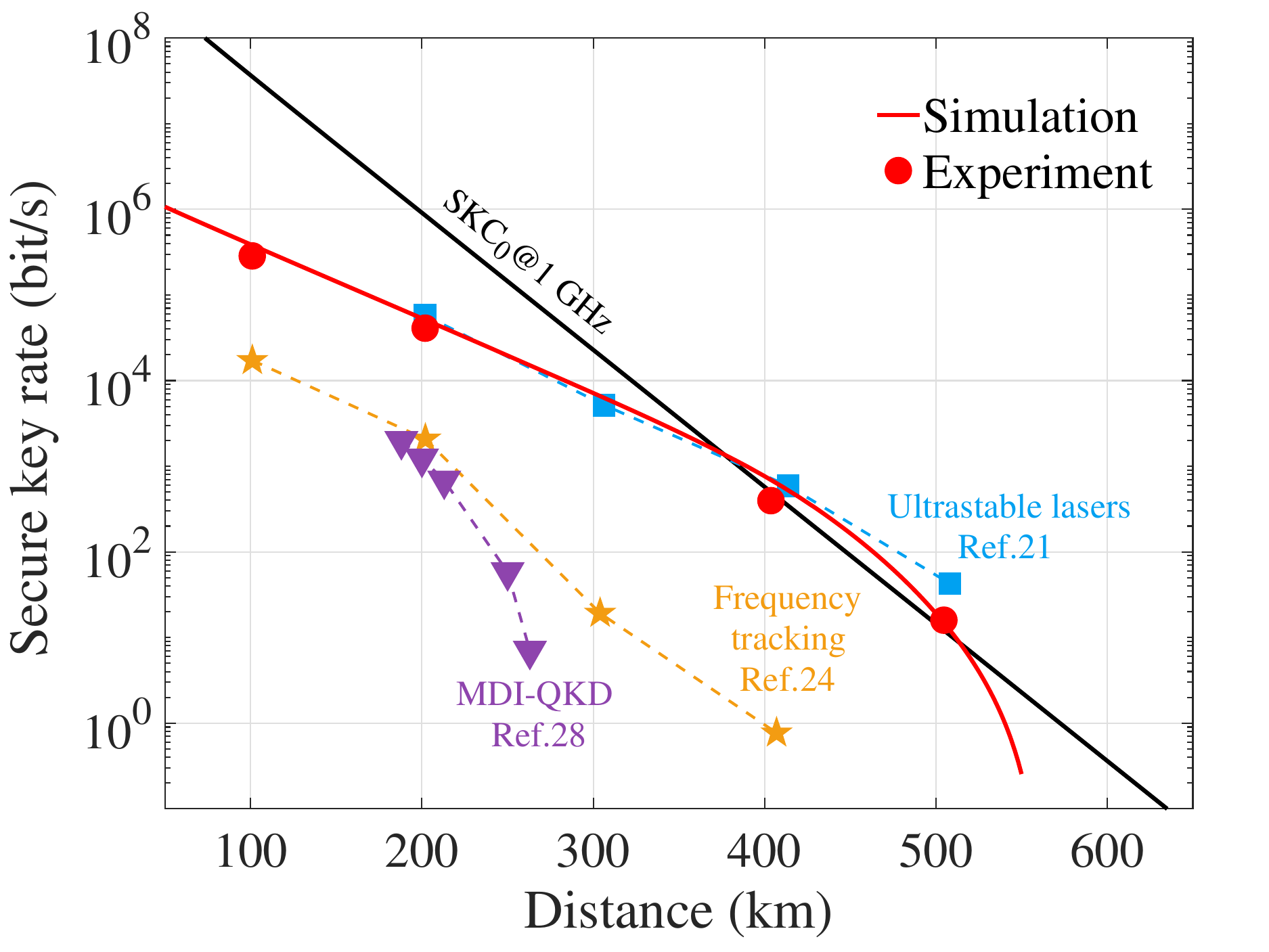}
	\caption{\textbf{Secure key rates.} The solid red circles indicate the experimental finite-size SKRs. The simulation (solid red line) and absolute repeaterless bound, $\rm{SKC}_0$ (solid black line) are plotted with 0.160~dB$/$km fiber attenuation for 1~GHz clock. In the SKR simulation, we set an empirical phase drift rate of 2.2~kHz, frequency offset of 260~Hz and a transmitted size of $2.54\times 10^{14}$  quantum pulses. 
  For comparison, we include PMP-QKD results using ultra-stable lasers~\cite{PhysRevLett.130.250801} as well as state-of-the-art MDI-QKD experiments using optical frequency tracking \zy{for reducing coincidence pairing error}~\cite{zhu2022experimental} \zy{or} pre-measurement coincidence pairing \zy{between just adjacent time-bins}~\cite{woodward2021gigahertz}.} \label{fig:SKR}
\end{figure}

PMP-QKD experiments are performed over different fiber distances, corresponding to channel losses from 16.98~dB (100.94~km) to 80.65~dB (504.66~km).  The details about the experimental parameters and results are summarized in Appendix~C.
We plot the finite-size SKRs obtained in experiments (solid red circles) and theoretical simulations (solid red line) in Fig~\ref{fig:SKR}. 
With finite-size effects taken into account, we record SKRs of 285.68k, 40.93k, 396.36 and 15.94~bit$/$s for 100.94, 201.88, 403.72 and 504.66~km, respectively.  The experimental results are in good agreement with the simulations. 

To visualize the repeater-like behavior of our  setup, we include in Fig.~\ref{fig:SKR} the absolute linear repeaterless bound (SKC$_0$)~\cite{pirandola17} for a point-to-point 1~GHz quantum link and thus highlight the square-root SKR scaling of our PMP-QKD setup.
Comparing to our previous results (blue squares) achieved with ultra-stable lasers~\cite{PhysRevLett.130.250801}, the present setup with compact, acetylene-stabilized lasers has a comparable performance for fiber distances  below 300~km but exhibits a noticeable performance drop at 400~km and above.  
We attribute this deterioration to the short-term phase instability of the compact lasers, as demonstrated earlier (see Fig.~\ref{FRE}).  Nevertheless, our setup is still able to overcome the absolute repeaterless bound by a margin of \lz{1.28}~times at 504.66~km. 

Figure~\ref{fig:SKR} compares our SKRs with recent results reported for other coincidence based QKD systems~\cite{woodward2021gigahertz,zhu2022experimental} using compact lasers. 
To counter against the drifting laser frequencies, these systems either adopt conventional \zy{adjacent time-bin coincidence pairing}~\cite{woodward2021gigahertz} for frequency  tolerance leading to a rapidly decreasing key rate, 
or implement optical frequency tracking~\cite{zhu2022experimental} which sacrifices a portion of transmission slots for frequency monitoring and hence reduces the SKRs. 
In contrast, our system uses the acetylene standard for long-term frequency stability and  delivers much higher SKRs with 100~\% quantum transmission duty cycle.

\textit{Discussion.}---We have demonstrated a practical post-measurement pairing QKD system using compact lasers referenced to acetylene cells as the local frequency standards.
Our data show that such standard enables excellent long-term frequency stability and allows an acceptable phase instability between lasers referenced to different cells. System performance over long fibers (400~km and above) can be improved through either increasing the system clock rate~\cite{wang22}, which will proportionally reduce the pairing interval and thus reduce the impact of the phase instability, or development of stabler locking to the absolute frequency standard. Our work brings forward an economical intercity quantum-secure network with absolute local frequency reference.

\vspace{5pt}

\textit{Note added.}---We note that related experimental work has been reported in Ref.~\cite{chen2024twin}. 
Both our work and Ref.~\cite{chen2024twin} implement the experiment with the local frequency standard. However, our system demonstrates the post-measurement coincidence pairing QKD and overcomes the repeaterless bound without global phase tracking, while Ref.~\cite{chen2024twin} demonstrates the twin-field QKD with strong phase reference pulse to estimate the relative phase difference.

 



\appendix
\begin{appendices}

\section*{APPENDIX A: LASER SOURCE CHARACTERIZATION}
\label{laser}

Two acetylene-stabilized lasers (Stabi$\lambda$ lasers $1542^\varepsilon$, Danish National Metrology Institute) are employed in the experiment. They exhibit narrow linewidth ($\leq 300$ Hz, in short term) and excellent long-term stability ($\leq 3 \times 10^{-13}$, Allan Deviation's sampling interval $\geq 1$ s). 
%
%
%
%
We measure the first-order interference between two continuous-wave, independently acetylene-stabilized lasers using a power meter over a duration of 100~s at a sampling rate of 500~kHz. The \zy{interference} visibility ($V$) is computed for \zy{every 10~ms} segment using the average ($\bar{I}_{max}$) of 10 highest values and the average ($\bar{I}_{min}$) of the 10 lowest values as,
\begin{equation}
V = \frac{\bar{I}_{max}-\bar{I}_{min}}{\bar{I}_{max}+\bar{I}_{min}}.
\end{equation}
The average visibility is 99.43~\% with a standard deviation of 0.50~\%.

\begin{figure}[ht]
\centering
\includegraphics[width=0.9\columnwidth]{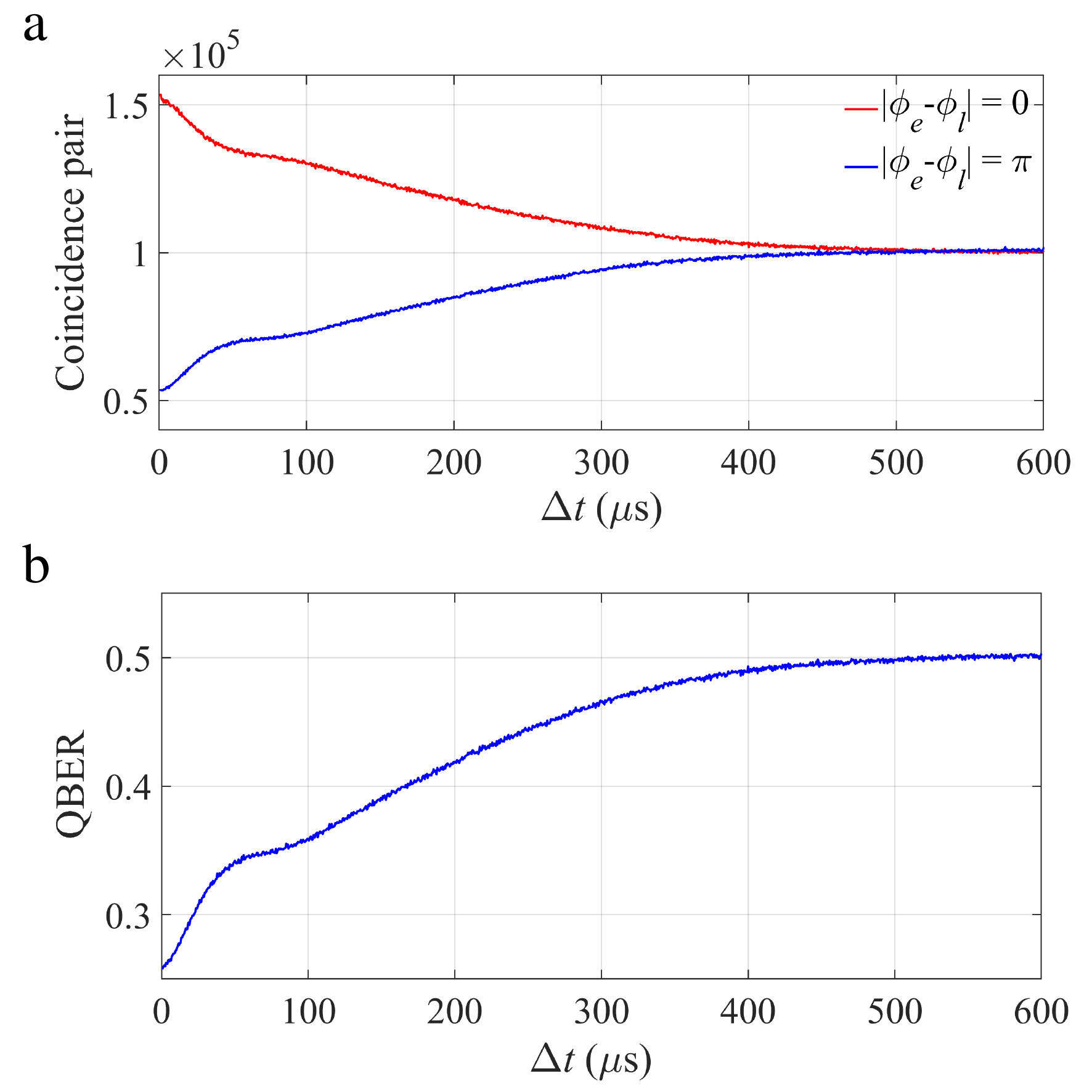}
\caption{\textbf{Characteristics of laser's stability and asynchronous two photons interference. } }
\label{freuqney difference}
\end{figure}

We then characterize the asynchronous two-photon interference 
using superconducting nanowire single-photon detectors (SNSPDs) using the setup shown in Fig.~1, Main Text. \zy{For time-bin econding,}  each laser's output is carved into pulses of 300~ps width at a clock frequency of 1~GHz.  Alice's pulses undergo alternative 0 and $\pi$ phase modulations while Bob's pulses \zy{are not} modulated. Alice and Bob's pulses interfere at Charlie's 50/50 beam splitter. 

We collect 10~s data and sort out the coincidences according to their temporal separation ($\Delta t = t_e - t_l$) and the encoded phase difference ($\Delta \phi = \phi_e - \phi_l$) between the early ($e$) and late ($l$) time-bins. 
In this analysis, it is sufficient to consider just photon 
clicks by one of the detectors. 
Figure~\ref{freuqney difference}\textbf{a} shows the number of coincidence counts as a function of $\Delta t$ for two different $\Delta \phi$ values. 
At short temporal delays, coincidences \lz{using the clicks in one detector} occur most frequently when $\Delta \phi = 0$ than $\Delta \phi = \pi$. \zy{As $\Delta t$ increases, the number of coincidences for $\Delta \phi = 0$ ($\Delta \phi = \pi$) decreases (increases) because of the lasers' frequency instability and hence the fluctuation of the differential phase between the two lasers.}
We calculate the quantum bit error rate (QBER) using  $QBER = \frac{C_{\pi}}{C_0 + C_\pi}$, where $C_0$ ($C_\pi$) represents the number of coincidences when the same detector clicks twice for $\Delta \phi = 0$ ($\Delta \phi = \pi$). In Fig.~\ref{freuqney difference}\textbf{b}, we observe a minimal QBER of 0.256 when $\Delta t \leq 0.5~\mu s$. The QBER  increases rapidly as $\Delta t$ increases to $50~\mu s$. The QBER saturates to 0.5 at $\Delta t = 500~\mu s$.



\section*{APPENDIX B: System loss and detector characterization}
\label{loss}

\begin{table}[ht]
\small
\setlength{\tabcolsep}{0.4pt}
\centering
\caption{Lengths and corresponding losses for the fiber links in the experiments.}
\centering
\begin{tabular}{c c|c c|c c}
\hline
\hline
\multicolumn{2}{c}{Total} & \multicolumn{2}{|c}{Alice} &  \multicolumn{2}{|c}{Bob}\\
\hline
length (km) & loss (dB) & length (km) & loss (dB) & length (km) & loss (dB)\\
\hline
\hline
100.94 & 16.98 & 50.47 & 8.64 & 50.47 & 8.34\\
201.88 & 32.47 & 100.94 & 16.45 & 100.94 & 16.02\\
403.72 & 64.01 & 201.87 & 32.14 & 201.85 & 31.87\\
504.66 & 80.65 & 252.34 & 40.5 & 252.32 & 40.15\\
\hline
\hline
\end{tabular}
\label{tab:fiber_los}
\end{table}

The quantum channel is established using ultra-low-loss fiber spools with average attenuation coefficients ranging from 0.159~dB~km$^{-1}$ to 0.168~dB~km$^{-1}$ in different distance links. The specific loss values for different distance links are presented in Table~\ref{tab:fiber_los}. In Charlie, there are a series of components including DWDM filters, polarization beam splitters, a 50:50 beam splitter, polarization controllers and fiber connectors. We summarize the loss values for these components in Table~\ref{tab:Components}.
Two SNSPDs are used to record the interference result of the quantum signal pulses. The characteristics of the detectors are summarized in Table~\ref{tab:Detectors}, with the dark count rates averaged over a measurement duration of 100 s.

\begin{table}[ht]
\setlength{\tabcolsep}{6pt}
\caption{Components loss in the Charlie.}
\centering

\begin{tabular}{c|c|c}
\hline
\hline
\makebox[0.1\textwidth][c]{} &  \makebox[0.1\textwidth][c]{Alice} &\makebox[0.1\textwidth][c]{Bob}  \\
\hline
DWDM filter & 0.68 & 0.78\\
\hline
Polarisation beam splitter & 0.38 & 0.58\\
\hline
50:50 beam splitter &0.81 &0.69 \\
\hline
Polarization controller &0.43 & 0.43\\
\hline
Connectors & 0.60 & 0.60\\
\hline
Total loss (dB)&2.90 & 3.08\\

\hline
\hline
\end{tabular}
\label{tab:Components}
\end{table}

\begin{table}[ht]
\setlength{\tabcolsep}{11pt}
\caption{Characteristics of Charlie's detectors $D_0$ and $D_1$ .}
\centering

\begin{tabular}{c|c|c}
\hline
\hline
\makebox[0.1\textwidth][c]{Detector} &  \makebox[0.1\textwidth][c]{Efficiency} &\makebox[0.15\textwidth][c]{Dark Count Rate}\\
\hline

$D_0$ & $71.0 \%$ & 6.3 Hz\\
\hline
$D_1$ & $70.5 \%$ & 9.0 Hz\\
\hline
\hline
\end{tabular}
\label{tab:Detectors}
\end{table}


\section*{APPENDIX C: Detailed experimental parameters and results}
\label{result}


\zy{Experimental setting parameters and results are summarized in Table~\ref{tab:key rate}.
The PMP-QKD protocol implements three intensities ($\mu$, $\nu$ and $o = 0$) with their corresponding \zy{sent} probabilities of $p_\mu$, $p_\nu$ and $p_o$.} 
Alice and Bob employ identical encoding parameters since all the experiments use symmetric fiber links. $F$ represents the clock frequency in system,  $N$  the total number of quantum signal pulses sent, \zy{and} $T_c$  the maximal pairing interval used in data post-processing. A successful click is denoted as $(k_{a}|k_{b})$ when Alice \lz{used} intensity $k_a$ and Bob $k_b$. $n_{[k_{a}^{tot},k_{b}^{tot}]}$ is the number of successful \lz{coincidences}, where $k_{a}^{tot}$ ($k_{a}^{tot}$) is the combined intensity of two time bins in Alice's (Bob's) side ($k_a^i,k_a^j \in \{\mu,\nu,o\}$, $k_b^i,k_b^j\in \{\mu,\nu,o\}$, $k_{a}^{tot}=k_a^i+k_a^j$, $k_{b}^{tot} = k_b^i+k_b^j$). $\overline{T}$ of $S_{[2\mu,2\mu]}$ and $S_{[2\nu,2\nu]}$ represent the average pairing intervals of $[2\mu_a,2\mu_b]$ and $[2\nu_a,2\nu_b]$ coincidences, respectively. The total numbers of error pairings in the $\boldsymbol{Z}$ and $\boldsymbol{X}$ bases are $m_{[\mu, \mu]}$ and $m_{[2\nu, 2\nu]}$, respectively. $E_z$ and $E_x$ indicate the error rates in the $\boldsymbol{Z}$ and $\boldsymbol{X}$ bases. The lower bounds of the number of single-photon pairs in the $\boldsymbol{Z}$ and $\boldsymbol{X}$ bases are denoted as $\underline{s}_{11}^{z}$ and $\underline{s}_{11}^{x}$, respectively. $\overline{\phi}_{11}^{z}$ represents the upper bound of the phase error rate for single-photon pairs in the $\boldsymbol{Z}$ basis. The SKR denotes the secure key rate, represented in different units such as per second and per clock. $SKC_0$ represents the absolute key capacity in a point-to-point link without the use of repeaters. The Ratio SKR over $SKC_0$ represents the ability to surpass the fundamental limit imposed by repeater-less communication.


\begin{table*}[ht]
\centering
\renewcommand{\arraystretch}{1.12}
\caption{Experimental parameters and results at various quantum link fiber lengths.}
\setlength{\tabcolsep}{3mm}{
\begin{tabular}{c | c c c c }
\hline \hline
    Total length (km) &   100.94  & 201.88  &403.72  & 504.66  \\
    \hline
$\mu$ & 0.431 &0.431 &0.424 & 0.542\\
$\nu$ & 0.020& 0.020&0.030 & 0.035\\
$p_{\mu}$ & 0.252& 0.252 & 0.217 & 0.261\\
$p_{\nu}$ &  0.194 & 0.194 & 0.315 & 0.344 \\
$p_{o}$ &  0.554 & 0.554 & 0.468 &  0.395\\
 \hline
  $F $ (Hz)   & $10^{9}$ & $10^{9}$ & $10^{9}$  &  $10^{9}$  \\ 
    $N$  & $2.133\times10^{12}$ & $5.04\times10^{12}$ & $7.96\times10^{13}$ &  $2.54\times10^{14}$ \\  
    $T_c$ ($\mu s$) & 3 & 5 & 60 & 77\\
    $\overline{T}$ of $\mathcal{S}_{[2\mu,2\mu]}$ ($\mu s$) & 0.113 & 0.651 & 20.830 & 35.228\\
     $\overline{T}$ of $\mathcal{S}_{[2\nu,2\nu]}$ ($\mu s$) & 0.114 & 0.652 & 20.800 & 35.250 \\
    \hline
  $n_{(\mu|\nu)}$ & 2283109970
 & 	928496635	 & 529763775	 & 409886657\\
  $n_{(\nu|\mu)}$ & 2296497980
 &  	965626544 & 	569429357 & 	433929248\\	 
	 \hline
    $n_{[o,o]}$ &1355 & 68 & 639 & 273 \\ 
    $n_{[\nu ,\nu] }$ & 2889713
	& 1164498 &	2331676 &  530309\\
    $n_{[\mu ,\mu] }$ & 2060950139
 &	843918892	&218758390 & 72092710\\
     $m_{[\mu ,\mu] }$ & 1987932  &	270424	& 157791 & 166221 \\
   $n_{[\nu ,o ]}$ & 69773 & 9719 & 17200 & 10460\\
   $n_{[\mu ,o] }$ & 1990057 &	272811 & 164716 & 123114\\
   $n_{[o ,\nu] }$ & 71380 &	10375 &	18979 & 11566\\
   $n_{[o , \mu ]}$ & 1995995 & 283279	& 173039 &132146\\ 
    $n_{[2\nu ,2\nu ]} $ &	81702 &	34009 &	267635 & 94966\\ 
   $m_{[2\nu ,2\nu] } $ &	21878  &	9248  &	 76634 & 29043 \\   
   $n_{[2\nu , o ]} $ & 1392098 &	540756 &	1044727 & 233897 \\ 
  $n_{[o , 2\nu] } $ & 1451838 &	621474 &	1273845 & 280885\\ 
  $n_{[2\mu , 2\mu] } $ & 106894915 &	44357293 &	11979746 & 7576903\\ 
  $m_{[2\mu , 2\mu] } $ & 27804718 &	11594290 &	3377479 &  2295000\\ 
 \hline
 $E_z$ &	0.00096 & 0.00032	 & 0.00072 & 0.00231\\
$E_x$ &	0.2678 &	0.2719	 & 0.2863 &  0.3058\\
 $\underline{s}_{11}^z$ & 903767681
& 371551409  & 88747013  &  21742779\\
 $\underline{s}_{11}^x$  & 38652  & 15892 & 109871 & 37571\\
$\overline{\phi}_{11}^{z}$ & 0.0531 & 0.0895 & 0.1551 & 0.2054
\\
 \hline
 SKR (bit/s)	 & $ 2.8568\times 10^{5}$
 & $4.0934\times 10^{4}$
 & $3.9636\times 10^{2}$
&$15.9350$
\\
SKR (bit/clock)	 & $2.8568\times 10^{-4}$
 & $4.0934\times 10^{-5}$
 & $3.9636\times 10^{-7}$
&$1.5935\times 10^{-8}$
\\
$SKC_{0}$  (bit/clock) &	$2.9212\times 10^{-2}$ & $8.1714\times 10^{-4}$& $5.7303\times 10^{-7}$ & $1.2422\times 10^{-8}$\\
Ratio SKR over $SKC_{0}$&	0.0098 &	0.0501	&   0.6917
  & 1.2828\\
\hline \hline
\end{tabular}
}
\label{tab:key rate}
\end{table*}
\centering

\end{appendices}









\newpage
\bigskip
This work was supported by the National Natural Science Foundation of China (62250710162, 62105034) and Beijing Natural Science Foundation (Z230005).




\bibliographystyle{modified-apsrev4-2_new}
\bibliography{AMDI_main_arxiv}

\begin{thebibliography}{31}%
\makeatletter
\providecommand \@ifxundefined [1]{%
 \@ifx{#1\undefined}
}%
\providecommand \@ifnum [1]{%
 \ifnum #1\expandafter \@firstoftwo
 \else \expandafter \@secondoftwo
 \fi
}%
\providecommand \@ifx [1]{%
 \ifx #1\expandafter \@firstoftwo
 \else \expandafter \@secondoftwo
 \fi
}%
\providecommand \natexlab [1]{#1}%
\providecommand \enquote  [1]{``#1''}%
\providecommand \bibnamefont  [1]{#1}%
\providecommand \bibfnamefont [1]{#1}%
\providecommand \citenamefont [1]{#1}%
\providecommand \href@noop [0]{\@secondoftwo}%
\providecommand \href [0]{\begingroup \@sanitize@url \@href}%
\providecommand \@href[1]{\@@startlink{#1}\@@href}%
\providecommand \@@href[1]{\endgroup#1\@@endlink}%
\providecommand \@sanitize@url [0]{\catcode `\\12\catcode `\$12\catcode `\&12\catcode `\#12\catcode `\^12\catcode `\_12\catcode `\%12\relax}%
\providecommand \@@startlink[1]{}%
\providecommand \@@endlink[0]{}%
\providecommand \url  [0]{\begingroup\@sanitize@url \@url }%
\providecommand \@url [1]{\endgroup\@href {#1}{\urlprefix }}%
\providecommand \urlprefix  [0]{URL }%
\providecommand \Eprint [0]{\href }%
\providecommand \doibase [0]{https://doi.org/}%
\providecommand \selectlanguage [0]{\@gobble}%
\providecommand \bibinfo  [0]{\@secondoftwo}%
\providecommand \bibfield  [0]{\@secondoftwo}%
\providecommand \translation [1]{[#1]}%
\providecommand \BibitemOpen [0]{}%
\providecommand \bibitemStop [0]{}%
\providecommand \bibitemNoStop [0]{.\EOS\space}%
\providecommand \EOS [0]{\spacefactor3000\relax}%
\providecommand \BibitemShut  [1]{\csname bibitem#1\endcsname}%
\let\auto@bib@innerbib\@empty
\bibitem [{\citenamefont {Dynes}\ \emph {et~al.}(2019)Dynes, J. F. and Wonfor, Adrian and Tam, W. W-S and Sharpe, A. W. and Takahashi, R. and Lucamarini, M. and Plews, A. and Yuan, Z. L. and Dixon, A. R. and Cho, J. and others}]{dynes2019cambridge}%
  \BibitemOpen
  \bibfield  {author} {\bibinfo {author} {\bibfnamefont {J.~F.}\ \bibnamefont {Dynes}}  \emph {et~al.},\ }\bibfield  {title} {\bibinfo {title} {{\color{Black}Cambridge quantum network}},\ }\href@noop {} {\bibfield  {journal} {\bibinfo  {journal} {npj Quant. Inf.}\ }\textbf {\bibinfo {volume} {5}},\ \bibinfo {pages} {101} (\bibinfo {year} {2019})}\BibitemShut {NoStop}%
\bibitem [{\citenamefont {Chen}\ \emph {et~al.}(2021)Chen, Yu-Ao and Zhang, Qiang and Chen, Teng-Yun and Cai, Wen-Qi and Liao, Sheng-Kai and Chen, Jun Kai and Yin, Juan and Ren, Ji-Gang and Chen, Zhu and Han, Sheng-Long and Yu, Qing and Liang, Ken and Zhou, Fei and Yuan, Xiao and Zhao, Mei-Sheng and Wang, Tian-Yin and Jiang, Xiao and Zhang, Liang and Liu, Wei-Yue and Li, Yang and Shen, Qi and Cao, Yuan and Lu, Chao-Yang and Shu, Rong and Wang, Jian-Yu and Li, Li and Liu, Nai-Le and Xu, Feihu and Wang, Xiang-Bin and Peng, Cheng-Zhi and Pan, Jian-Wei}]{chen2021integrated}%
  \BibitemOpen
  \bibfield  {author} {\bibinfo {author} {\bibfnamefont {Y.-A.}\ \bibnamefont {Chen}}  \emph {et~al.},\ }\bibfield  {title} {\bibinfo {title} {{\color{Black}An integrated space-to-ground quantum communication network over 4,600 kilometres}},\ }\href@noop {} {\bibfield  {journal} {\bibinfo  {journal} {Nature}\ }\textbf {\bibinfo {volume} {589}},\ \bibinfo {pages} {214} (\bibinfo {year} {2021})}\BibitemShut {NoStop}%
\bibitem [{\citenamefont {Yuan}\ \emph {et~al.}(2018)Yuan, Zhiliang and Plews, Alan and Takahashi, Ririka and Doi, Kazuaki and Tam, Winci and Sharpe, Andrew W and Dixon, Alexander R and Lavelle, Evan and Dynes, James F and Murakami, Akira and others}]{yuan201810}%
  \BibitemOpen
  \bibfield  {author} {\bibinfo {author} {\bibfnamefont {Z.}~\bibnamefont {Yuan}}  \emph {et~al.},\ }\bibfield  {title} {\bibinfo {title} {{\color{Black}10-\uppercase{M}b/s quantum key distribution}},\ }\href@noop {} {\bibfield  {journal} {\bibinfo  {journal} {J. Lightwave Technol.}\ }\textbf {\bibinfo {volume} {36}},\ \bibinfo {pages} {3427} (\bibinfo {year} {2018})}\BibitemShut {NoStop}%
\bibitem [{\citenamefont {Li}\ \emph {et~al.}(2023)Li, Wei and Zhang, Likang and Tan, Hao and Lu, Yichen and Liao, Sheng-Kai and Huang, Jia and Li, Hao and Wang, Zhen and Mao, Hao-Kun and Yan, Bingze and others}]{li2023high}%
  \BibitemOpen
  \bibfield  {author} {\bibinfo {author} {\bibfnamefont {W.}~\bibnamefont {Li}}  \emph {et~al.},\ }\bibfield  {title} {\bibinfo {title} {{\color{Black}High-rate quantum key distribution exceeding 110 {Mb} s$^{-1}$}},\ }\href@noop {} {\bibfield  {journal} {\bibinfo  {journal} {Nat. Photonics}\ }\textbf {\bibinfo {volume} {17}},\ \bibinfo {pages} {416} (\bibinfo {year} {2023})}\BibitemShut {NoStop}%
\bibitem [{\citenamefont {Boaron}\ \emph {et~al.}(2018)Boaron, Alberto and Boso, Gianluca and Rusca, Davide and Vulliez, C\'edric and Autebert, Claire and Caloz, Misael and Perrenoud, Matthieu and Gras, Ga\"etan and Bussi\`eres, F\'elix and Li, Ming-Jun and Nolan, Daniel and Martin, Anthony and Zbinden, Hugo}]{PhysRevLett.121.190502}%
  \BibitemOpen
  \bibfield  {author} {\bibinfo {author} {\bibfnamefont {A.}~\bibnamefont {Boaron}}  \emph {et~al.},\ }\bibfield  {title} {\bibinfo {title} {{\color{Black}Secure quantum key distribution over 421 km of optical fiber}},\ }\href {https://doi.org/10.1103/PhysRevLett.121.190502} {\bibfield  {journal} {\bibinfo  {journal} {Phys. Rev. Lett.}\ }\textbf {\bibinfo {volume} {121}},\ \bibinfo {pages} {190502} (\bibinfo {year} {2018})}\BibitemShut {NoStop}%
\bibitem [{\citenamefont {Paraiso}\ \emph {et~al.}(2021)Paraiso, Taofiq K and Roger, Thomas and Marangon, Davide G and De Marco, Innocenzo and Sanzaro, Mirko and Woodward, Robert I and Dynes, James F and Yuan, Zhiliang and Shields, Andrew J}]{paraiso2021photonic}%
  \BibitemOpen
  \bibfield  {author} {\bibinfo {author} {\bibfnamefont {T.~K.}\ \bibnamefont {Paraiso}}  \emph {et~al.},\ }\bibfield  {title} {\bibinfo {title} {{\color{Black}A photonic integrated quantum secure communication system}},\ }\href@noop {} {\bibfield  {journal} {\bibinfo  {journal} {Nat. Photonics}\ }\textbf {\bibinfo {volume} {15}},\ \bibinfo {pages} {850} (\bibinfo {year} {2021})}\BibitemShut {NoStop}%
\bibitem [{\citenamefont {Sax}\ \emph {et~al.}(2023)Sax, Rebecka and Boaron, Alberto and Boso, Gianluca and Atzeni, Simone and Crespi, Andrea and Gr{\"u}nenfelder, Fadri and Rusca, Davide and Al-Saadi, Aws and Bronzi, Danilo and Kupijai, Sebastian and others}]{sax2023high}%
  \BibitemOpen
  \bibfield  {author} {\bibinfo {author} {\bibfnamefont {R.}~\bibnamefont {Sax}}  \emph {et~al.},\ }\bibfield  {title} {\bibinfo {title} {{\color{Black}High-speed integrated QKD system}},\ }\href@noop {} {\bibfield  {journal} {\bibinfo  {journal} {Photonics Res.}\ }\textbf {\bibinfo {volume} {11}},\ \bibinfo {pages} {1007} (\bibinfo {year} {2023})}\BibitemShut {NoStop}%
\bibitem [{\citenamefont {Xu}\ \emph {et~al.}(2020)Xu, Feihu and Ma, Xiongfeng and Zhang, Qiang and Lo, Hoi-Kwong and Pan, Jian-Wei}]{xu2020secure}%
  \BibitemOpen
  \bibfield  {author} {\bibinfo {author} {\bibfnamefont {F.}~\bibnamefont {Xu}}, \bibinfo {author} {\bibfnamefont {X.}~\bibnamefont {Ma}}, \bibinfo {author} {\bibfnamefont {Q.}~\bibnamefont {Zhang}}, \bibinfo {author} {\bibfnamefont {H.-K.}\ \bibnamefont {Lo}} \ and\ \bibinfo {author} {\bibfnamefont {J.-W.}\ \bibnamefont {Pan}},\ }\bibfield  {title} {\bibinfo {title} {{\color{Black}Secure quantum key distribution with realistic devices}},\ }\href@noop {} {\bibfield  {journal} {\bibinfo  {journal} {Rev. Mod. Phys.}\ }\textbf {\bibinfo {volume} {92}},\ \bibinfo {pages} {025002} (\bibinfo {year} {2020})}\BibitemShut {NoStop}%
\bibitem [{\citenamefont {Zapatero}\ \emph {et~al.}(2024)Zapatero, V{\'\i}ctor and Navarrete, {\'A}lvaro and Curty, Marcos}]{zapatero2024implementation}%
  \BibitemOpen
  \bibfield  {author} {\bibinfo {author} {\bibfnamefont {V.}~\bibnamefont {Zapatero}}, \bibinfo {author} {\bibfnamefont {{\'A}.}~\bibnamefont {Navarrete}} \ and\ \bibinfo {author} {\bibfnamefont {M.}~\bibnamefont {Curty}},\ }\bibfield  {title} {\bibinfo {title} {{\color{Black}Implementation security in quantum key distribution}},\ }\href@noop {} {\bibfield  {journal} {\bibinfo  {journal} {Advanced Quantum Technologies}\ ,\ \bibinfo {pages} {2300380}} (\bibinfo {year} {2024})}\BibitemShut {NoStop}%
\bibitem [{\citenamefont {Pirandola}\ \emph {et~al.}(2017)Pirandola, Stefano and Laurenza, Riccardo and Ottaviani, Carlo and Banchi, Leonardo}]{pirandola17}%
  \BibitemOpen
  \bibfield  {author} {\bibinfo {author} {\bibfnamefont {S.}~\bibnamefont {Pirandola}}, \bibinfo {author} {\bibfnamefont {R.}~\bibnamefont {Laurenza}}, \bibinfo {author} {\bibfnamefont {C.}~\bibnamefont {Ottaviani}} \ and\ \bibinfo {author} {\bibfnamefont {L.}~\bibnamefont {Banchi}},\ }\bibfield  {title} {\bibinfo {title} {{\color{Black}Fundamental limits of repeaterless quantum communications}},\ }\href@noop {} {\bibfield  {journal} {\bibinfo  {journal} {Nat. Commun.}\ }\textbf {\bibinfo {volume} {8}},\ \bibinfo {pages} {15043} (\bibinfo {year} {2017})}\BibitemShut {NoStop}%
\bibitem [{\citenamefont {Briegel}\ \emph {et~al.}(1998)Briegel, H-J and D{\"u}r, Wolfgang and Cirac, Juan I and Zoller, Peter}]{briegel1998quantum}%
  \BibitemOpen
  \bibfield  {author} {\bibinfo {author} {\bibfnamefont {H.-J.}\ \bibnamefont {Briegel}}, \bibinfo {author} {\bibfnamefont {W.}~\bibnamefont {D{\"u}r}}, \bibinfo {author} {\bibfnamefont {J.~I.}\ \bibnamefont {Cirac}} \ and\ \bibinfo {author} {\bibfnamefont {P.}~\bibnamefont {Zoller}},\ }\bibfield  {title} {\bibinfo {title} {{\color{Black}Quantum repeaters: the role of imperfect local operations in quantum communication}},\ }\href@noop {} {\bibfield  {journal} {\bibinfo  {journal} {Phys. Rev. Lett.}\ }\textbf {\bibinfo {volume} {81}},\ \bibinfo {pages} {5932} (\bibinfo {year} {1998})}\BibitemShut {NoStop}%
\bibitem [{\citenamefont {Duan}\ \emph {et~al.}(2001)Duan, L-M and Lukin, Mikhail D and Cirac, J Ignacio and Zoller, Peter}]{duan2001long}%
  \BibitemOpen
  \bibfield  {author} {\bibinfo {author} {\bibfnamefont {L.-M.}\ \bibnamefont {Duan}}, \bibinfo {author} {\bibfnamefont {M.~D.}\ \bibnamefont {Lukin}}, \bibinfo {author} {\bibfnamefont {J.~I.}\ \bibnamefont {Cirac}} \ and\ \bibinfo {author} {\bibfnamefont {P.}~\bibnamefont {Zoller}},\ }\bibfield  {title} {\bibinfo {title} {{\color{Black}Long-distance quantum communication with atomic ensembles and linear optics}},\ }\href@noop {} {\bibfield  {journal} {\bibinfo  {journal} {Nature}\ }\textbf {\bibinfo {volume} {414}},\ \bibinfo {pages} {413} (\bibinfo {year} {2001})}\BibitemShut {NoStop}%
\bibitem [{\citenamefont {van Leent}\ \emph {et~al.}(2022)van Leent, Tim and Bock, Matthias and Fertig, Florian and Garthoff, Robert and Eppelt, Sebastian and Zhou, Yiru and Malik, Pooja and Seubert, Matthias and Bauer, Tobias and Rosenfeld, Wenjamin and others}]{van2022entangling}%
  \BibitemOpen
  \bibfield  {author} {\bibinfo {author} {\bibfnamefont {T.}~\bibnamefont {van Leent}}  \emph {et~al.},\ }\bibfield  {title} {\bibinfo {title} {{\color{Black}Entangling single atoms over 33 km telecom fibre}},\ }\href@noop {} {\bibfield  {journal} {\bibinfo  {journal} {Nature}\ }\textbf {\bibinfo {volume} {607}},\ \bibinfo {pages} {69} (\bibinfo {year} {2022})}\BibitemShut {NoStop}%
\bibitem [{\citenamefont {Lucamarini}\ \emph {et~al.}(2018)Lucamarini, Marco and Yuan, Zhiliang L and Dynes, James F and Shields, Andrew J}]{Lucamarini2018}%
  \BibitemOpen
  \bibfield  {author} {\bibinfo {author} {\bibfnamefont {M.}~\bibnamefont {Lucamarini}}, \bibinfo {author} {\bibfnamefont {Z.~L.}\ \bibnamefont {Yuan}}, \bibinfo {author} {\bibfnamefont {J.~F.}\ \bibnamefont {Dynes}} \ and\ \bibinfo {author} {\bibfnamefont {A.~J.}\ \bibnamefont {Shields}},\ }\bibfield  {title} {\bibinfo {title} {{\color{Black}Overcoming the rate--distance limit of quantum key distribution without quantum repeaters}},\ }\href@noop {} {\bibfield  {journal} {\bibinfo  {journal} {Nature}\ }\textbf {\bibinfo {volume} {557}},\ \bibinfo {pages} {400} (\bibinfo {year} {2018})}\BibitemShut {NoStop}%
\bibitem [{\citenamefont {Zeng}\ \emph {et~al.}(2022)Zeng, Pei and Zhou, Hongyi and Wu, Weijie and Ma, Xiongfeng}]{zeng2022mode}%
  \BibitemOpen
  \bibfield  {author} {\bibinfo {author} {\bibfnamefont {P.}~\bibnamefont {Zeng}}, \bibinfo {author} {\bibfnamefont {H.}~\bibnamefont {Zhou}}, \bibinfo {author} {\bibfnamefont {W.}~\bibnamefont {Wu}} \ and\ \bibinfo {author} {\bibfnamefont {X.}~\bibnamefont {Ma}},\ }\bibfield  {title} {\bibinfo {title} {{\color{Black}Mode-pairing quantum key distribution}},\ }\href@noop {} {\bibfield  {journal} {\bibinfo  {journal} {Nat. Commun.}\ }\textbf {\bibinfo {volume} {13}},\ \bibinfo {pages} {3903} (\bibinfo {year} {2022})}\BibitemShut {NoStop}%
\bibitem [{\citenamefont {Xie}\ \emph {et~al.}(2022)Xie, Yuan-Mei and Lu, Yu-Shuo and Weng, Chen-Xun and Cao, Xiao-Yu and Jia, Zhao-Ying and Bao, Yu and Wang, Yang and Fu, Yao and Yin, Hua-Lei and Chen, Zeng-Bing}]{xie2022breaking}%
  \BibitemOpen
  \bibfield  {author} {\bibinfo {author} {\bibfnamefont {Y.-M.}\ \bibnamefont {Xie}}  \emph {et~al.},\ }\bibfield  {title} {\bibinfo {title} {{\color{Black}Breaking the rate-loss bound of quantum key distribution with asynchronous two-photon interference}},\ }\href@noop {} {\bibfield  {journal} {\bibinfo  {journal} {PRX Quantum}\ }\textbf {\bibinfo {volume} {3}},\ \bibinfo {pages} {020315} (\bibinfo {year} {2022})}\BibitemShut {NoStop}%
\bibitem [{\citenamefont {Pittaluga}\ \emph {et~al.}(2021)Pittaluga, Mirko and Minder, Mariella and Lucamarini, Marco and Sanzaro, Mirko and Woodward, Robert I. and Li, Ming-Jun and Yuan, Zhiliang and Shields, Andrew J.}]{pittaluga21}%
  \BibitemOpen
  \bibfield  {author} {\bibinfo {author} {\bibfnamefont {M.}~\bibnamefont {Pittaluga}}  \emph {et~al.},\ }\bibfield  {title} {\bibinfo {title} {{\color{Black}600-km repeater-like quantum communications with dual-band stabilization}},\ }\href@noop {} {\bibfield  {journal} {\bibinfo  {journal} {Nat. Photonics}\ }\textbf {\bibinfo {volume} {15}},\ \bibinfo {pages} {530} (\bibinfo {year} {2021})}\BibitemShut {NoStop}%
\bibitem [{\citenamefont {Wang}\ \emph {et~al.}(2022)Wang, Shuang and Yin, Zhen-Qian and He, De-Yong and Chen, Wei and Wang, Rui-Qiang and Ye, Peng and Zhou, Yao and Fan-Yuan, Guan-Jie and Wang, Fang-Xiang and Chen, Wei and Zhu, Yong-Gang and V. Morozov, Pavel and V. Divochiy, Alexander and Zhou, Zheng and Guo, Guang-Can and Han, Zheng-Fu}]{wang22}%
  \BibitemOpen
  \bibfield  {author} {\bibinfo {author} {\bibfnamefont {S.}~\bibnamefont {Wang}}  \emph {et~al.},\ }\bibfield  {title} {\bibinfo {title} {{\color{Black}Twin-field quantum key distribution over 830-km fibre}},\ }\href@noop {} {\bibfield  {journal} {\bibinfo  {journal} {Nat. Photonics}\ }\textbf {\bibinfo {volume} {16}},\ \bibinfo {pages} {154} (\bibinfo {year} {2022})}\BibitemShut {NoStop}%
\bibitem [{\citenamefont {Zhou}\ \emph {et~al.}(2023{\natexlab{a}})Zhou, Lai and Lin, Jinping and Jing, Yumang and Yuan, Zhiliang}]{zhou2023quantum}%
  \BibitemOpen
  \bibfield  {author} {\bibinfo {author} {\bibfnamefont {L.}~\bibnamefont {Zhou}}, \bibinfo {author} {\bibfnamefont {J.}~\bibnamefont {Lin}}, \bibinfo {author} {\bibfnamefont {Y.}~\bibnamefont {Jing}} \ and\ \bibinfo {author} {\bibfnamefont {Z.}~\bibnamefont {Yuan}},\ }\bibfield  {title} {\bibinfo {title} {{\color{Black}Twin-field quantum key distribution without optical frequency dissemination}},\ }\href@noop {} {\bibfield  {journal} {\bibinfo  {journal} {Nat. Commun.}\ }\textbf {\bibinfo {volume} {14}},\ \bibinfo {pages} {928} (\bibinfo {year} {2023}{\natexlab{a}})}\BibitemShut {NoStop}%
\bibitem [{\citenamefont {Zhong}\ \emph {et~al.}(2021)Zhong, Xiaoqing and Wang, Wenyuan and Qian, Li and Lo, Hoi-Kwong}]{zhong2021proof}%
  \BibitemOpen
  \bibfield  {author} {\bibinfo {author} {\bibfnamefont {X.}~\bibnamefont {Zhong}}, \bibinfo {author} {\bibfnamefont {W.}~\bibnamefont {Wang}}, \bibinfo {author} {\bibfnamefont {L.}~\bibnamefont {Qian}} \ and\ \bibinfo {author} {\bibfnamefont {H.-K.}\ \bibnamefont {Lo}},\ }\bibfield  {title} {\bibinfo {title} {{\color{Black}Proof-of-principle experimental demonstration of twin-field quantum key distribution over optical channels with asymmetric losses}},\ }\href@noop {} {\bibfield  {journal} {\bibinfo  {journal} {npj Quant. Inf.}\ }\textbf {\bibinfo {volume} {7}},\ \bibinfo {pages} {8} (\bibinfo {year} {2021})}\BibitemShut {NoStop}%
\bibitem [{\citenamefont {Zhou}\ \emph {et~al.}(2023{\natexlab{b}})Zhou, Lai and Lin, Jinping and Xie, Yuan-Mei and Lu, Yu-Shuo and Jing, Yumang and Yin, Hua-Lei and Yuan, Zhiliang}]{PhysRevLett.130.250801}%
  \BibitemOpen
  \bibfield  {author} {\bibinfo {author} {\bibfnamefont {L.}~\bibnamefont {Zhou}}  \emph {et~al.},\ }\bibfield  {title} {\bibinfo {title} {{\color{Black}Experimental quantum communication overcomes the rate-loss limit without global phase tracking}},\ }\href {https://doi.org/10.1103/PhysRevLett.130.250801} {\bibfield  {journal} {\bibinfo  {journal} {Phys. Rev. Lett.}\ }\textbf {\bibinfo {volume} {130}},\ \bibinfo {pages} {250801} (\bibinfo {year} {2023}{\natexlab{b}})}\BibitemShut {NoStop}%
\bibitem [{\citenamefont {Liu}\ \emph {et~al.}(2023)Liu, Yang and Zhang, Wei-Jun and Jiang, Cong and Chen, Jiu-Peng and Zhang, Chi and Pan, Wen-Xin and Ma, Di and Dong, Hao and Xiong, Jia-Min and Zhang, Cheng-Jun and others}]{liu2023experimental}%
  \BibitemOpen
  \bibfield  {author} {\bibinfo {author} {\bibfnamefont {Y.}~\bibnamefont {Liu}}  \emph {et~al.},\ }\bibfield  {title} {\bibinfo {title} {{\color{Black}Experimental twin-field quantum key distribution over 1000 km fiber distance}},\ }\href@noop {} {\bibfield  {journal} {\bibinfo  {journal} {Phys. Rev. Lett.}\ }\textbf {\bibinfo {volume} {130}},\ \bibinfo {pages} {210801} (\bibinfo {year} {2023})}\BibitemShut {NoStop}%
\bibitem [{\citenamefont {Avesani}(2023)Avesani, Marco}]{avesani2023long}%
  \BibitemOpen
  \bibfield  {author} {\bibinfo {author} {\bibfnamefont {M.}~\bibnamefont {Avesani}},\ }\bibfield  {title} {\bibinfo {title} {{\color{Black}Long-Range Quantum Cryptography Gets Simpler}},\ }\href@noop {} {\bibfield  {journal} {\bibinfo  {journal} {Physics}\ }\textbf {\bibinfo {volume} {16}},\ \bibinfo {pages} {104} (\bibinfo {year} {2023})}\BibitemShut {NoStop}%
\bibitem [{\citenamefont {Zhu}\ \emph {et~al.}(2023)Zhu, Hao-Tao and Huang, Yizhi and Liu, Hui and Zeng, Pei and Zou, Mi and Dai, Yunqi and Tang, Shibiao and Li, Hao and You, Lixing and Wang, Zhen and others}]{zhu2022experimental}%
  \BibitemOpen
  \bibfield  {author} {\bibinfo {author} {\bibfnamefont {H.-T.}\ \bibnamefont {Zhu}}  \emph {et~al.},\ }\bibfield  {title} {\bibinfo {title} {{\color{Black}Experimental mode-pairing measurement-device-independent quantum key distribution without global phase locking}},\ }\href@noop {} {\bibfield  {journal} {\bibinfo  {journal} {Phys. Rev. Lett.}\ }\textbf {\bibinfo {volume} {130}},\ \bibinfo {pages} {030801} (\bibinfo {year} {2023})}\BibitemShut {NoStop}%
\bibitem [{\citenamefont {Wei}\ \emph {et~al.}(2020)Wei, Kejin and Li, Wei and Tan, Hao and Li, Yang and Min, Hao and Zhang, Wei-Jun and Li, Hao and You, Lixing and Wang, Zhen and Jiang, Xiao and Chen, Teng-Yun and Liao, Sheng-Kai and Peng, Cheng-Zhi and Xu, Feihu and Pan, Jian-Wei}]{wei2020high}%
  \BibitemOpen
  \bibfield  {author} {\bibinfo {author} {\bibfnamefont {K.}~\bibnamefont {Wei}}  \emph {et~al.},\ }\bibfield  {title} {\bibinfo {title} {{\color{Black}High-speed measurement-device-independent quantum key distribution with integrated silicon photonics}},\ }\href@noop {} {\bibfield  {journal} {\bibinfo  {journal} {Phys. Rev. X}\ }\textbf {\bibinfo {volume} {10}},\ \bibinfo {pages} {031030} (\bibinfo {year} {2020})}\BibitemShut {NoStop}%
\bibitem [{\citenamefont {des Poids~et Mesures~(BIPM)}()Bureau Internatinal des Poids et Mesures (BIPM)}]{fre_standard}%
  \BibitemOpen
  \bibfield  {author} {\bibinfo {author} {\bibfnamefont {B.~I.}\ \bibnamefont {des Poids~et Mesures~(BIPM)}},\ }\href@noop {} {\bibinfo {title} {{\color{Black}Recommended values of standard frequencies}}},\ \bibinfo {note} {see [retrieved 20 May 2024] \url{https://www.bipm.org/documents/20126/41590985/M-e-P_13C2H2_1.54.pdf/623e90e2-e816-36f7-2fd3-93b2e741ee5c}}\BibitemShut {NoStop}%
\bibitem [{\citenamefont {Lo}\ \emph {et~al.}(2012)Lo, Hoi-Kwong and Curty, Marcos and Qi, Bing}]{lo2012measurement}%
  \BibitemOpen
  \bibfield  {author} {\bibinfo {author} {\bibfnamefont {H.-K.}\ \bibnamefont {Lo}}, \bibinfo {author} {\bibfnamefont {M.}~\bibnamefont {Curty}} \ and\ \bibinfo {author} {\bibfnamefont {B.}~\bibnamefont {Qi}},\ }\bibfield  {title} {\bibinfo {title} {{\color{Black}Measurement-device-independent quantum key distribution}},\ }\href@noop {} {\bibfield  {journal} {\bibinfo  {journal} {Phys. Rev. Lett.}\ }\textbf {\bibinfo {volume} {108}},\ \bibinfo {pages} {130503} (\bibinfo {year} {2012})}\BibitemShut {NoStop}%
\bibitem [{\citenamefont {Woodward}\ \emph {et~al.}(2021)Woodward, Robert Ian and Lo, YS and Pittaluga, M and Minder, M and Para{\"\i}so, TK and Lucamarini, M and Yuan, ZL and Shields, AJ}]{woodward2021gigahertz}%
  \BibitemOpen
  \bibfield  {author} {\bibinfo {author} {\bibfnamefont {R.~I.}\ \bibnamefont {Woodward}}  \emph {et~al.},\ }\bibfield  {title} {\bibinfo {title} {{\color{Black}Gigahertz measurement-device-independent quantum key distribution using directly modulated lasers}},\ }\href@noop {} {\bibfield  {journal} {\bibinfo  {journal} {npj Quant. Inf.}\ }\textbf {\bibinfo {volume} {7}},\ \bibinfo {pages} {58} (\bibinfo {year} {2021})}\BibitemShut {NoStop}%
\bibitem [{\citenamefont {Hald}\ \emph {et~al.}(2011)Hald, Jan and Nielsen, Lars and Petersen, Jan C and Varming, Poul and Pedersen, Jens E}]{hald2011fiber}%
  \BibitemOpen
  \bibfield  {author} {\bibinfo {author} {\bibfnamefont {J.}~\bibnamefont {Hald}}, \bibinfo {author} {\bibfnamefont {L.}~\bibnamefont {Nielsen}}, \bibinfo {author} {\bibfnamefont {J.~C.}\ \bibnamefont {Petersen}}, \bibinfo {author} {\bibfnamefont {P.}~\bibnamefont {Varming}} \ and\ \bibinfo {author} {\bibfnamefont {J.~E.}\ \bibnamefont {Pedersen}},\ }\bibfield  {title} {\bibinfo {title} {{\color{Black}Fiber laser optical frequency standard at 1.54 $\mu$m}},\ }\href@noop {} {\bibfield  {journal} {\bibinfo  {journal} {Opt. Express}\ }\textbf {\bibinfo {volume} {19}},\ \bibinfo {pages} {2052} (\bibinfo {year} {2011})}\BibitemShut {NoStop}%
\bibitem [{\citenamefont {Talvard}\ \emph {et~al.}(2017)Thomas Talvard and Philip G. Westergaard and Michael V. DePalatis and Nicolai F. Mortensen and Michael Drewsen and Bjarke G{\o}th and Jan Hald}]{Talvard:17}%
  \BibitemOpen
  \bibfield  {author} {\bibinfo {author} {\bibfnamefont {T.}~\bibnamefont {Talvard}}  \emph {et~al.},\ }\bibfield  {title} {\bibinfo {title} {{\color{Black}Enhancement of the performance of a fiber-based frequency comb by referencing to an acetylene-stabilized fiber laser}},\ }\href {https://doi.org/10.1364/OE.25.002259} {\bibfield  {journal} {\bibinfo  {journal} {Opt. Express}\ }\textbf {\bibinfo {volume} {25}},\ \bibinfo {pages} {2259} (\bibinfo {year} {2017})}\BibitemShut {NoStop}%
\bibitem [{\citenamefont {Chen}\ \emph {et~al.}(2024)Chen, Jiu-Peng and Zhou, Fei and Zhang, Chi and Jiang, Cong and Chen, Fa-Xi and Huang, Jia and Li, Hao and You, Li-Xing and Wang, Xiang-Bin and Liu, Yang and others}]{chen2024twin}%
  \BibitemOpen
  \bibfield  {author} {\bibinfo {author} {\bibfnamefont {J.-P.}\ \bibnamefont {Chen}}  \emph {et~al.},\ }\bibfield  {title} {\bibinfo {title} {{\color{Black}Twin-field quantum key distribution with local frequency reference}},\ }\href@noop {} {\bibfield  {journal} {\bibinfo  {journal} {Physical Review Letters}\ }\textbf {\bibinfo {volume} {132}},\ \bibinfo {pages} {260802} (\bibinfo {year} {2024})}\BibitemShut {NoStop}%
\end{thebibliography}%



\end{document}